\documentclass[twocolumn,prd,amssymb,nofootinbib,balancelastpage,amsmath,preprintnumbers]{revtex4}
\usepackage{graphicx}
\usepackage{dcolumn}
\usepackage{bm}
\usepackage{verbatim}

\def\lsim{\mathrel{\mathstrut\smash{\ooalign{\raise2.5pt\hbox{$<$}\cr\lower2.5pt\hbox{$\sim$}}}}}
\def\gsim{\mathrel{\mathstrut\smash{\ooalign{\raise2.5pt\hbox{$>$}\cr\lower2.5pt\hbox{$\sim$}}}}}

\def\be{\begin{equation}}
\def\ee{\end{equation}}
\def\bea{\begin{eqnarray}}
\def\eea{\end{eqnarray}}

\def\sigv{\langle \sigma v \rangle}
\def\ave#1{\langle #1 \rangle}

\begin{document}

\title{ Cosmological Limits on Hidden Sector Dark Matter}

\author{Subinoy Das}
\email{subinoy@physics.ubc.ca}
\affiliation{Department of Physics and Astronomy, University of British Columbia, Vancouver, BC V6T 1Z1, Canada}

\author{Kris Sigurdson}
\email{krs@physics.ubc.ca}
\affiliation{Department of Physics and Astronomy, University of British Columbia, Vancouver, BC V6T 1Z1, Canada}

\begin{abstract} 
We explore the model-independent constraints from cosmology on a dark-matter particle with no prominent standard model interactions  that interacts and thermalizes with other particles in a hidden sector.  Without specifying detailed hidden-sector particle physics, we characterize the relevant physics by the annihilation cross section, mass,  and temperature ratio of the hidden to visible sectors.  
While encompassing the standard cold WIMP scenario, we do not require the freeze-out process to be nonrelativistic. Rather, freeze-out may also occur when dark matter particles are semirelativistic or relativistic.  We solve the Boltzmann equation 
to find the conditions that hidden-sector dark matter accounts for the observed dark-matter density, satisfies the Tremaine-Gunn bound on dark-matter phase space density, and has a free-streaming length consistent with cosmological constraints on the matter power spectrum.  We show that for masses $\lesssim\! 1.5$~keV no region of parameter space satisfies all these constraints. This is a gravitationally-mediated  lower bound on the dark-matter mass for any model in which the primary component of daqrk matter once had efficient interactions --- even if it has never been in equilibrium with the standard model.
 \end{abstract}
\maketitle 

\section{ Introduction}

For many decades we have been aware that dark matter exists through its gravitational effects on galaxies and the Universe as a whole.\footnote{ See, e.g., Ref.~\cite{D'Amico:2009df} for a pedagogical review.}  
Over this time, compelling candidates for dark matter have been proposed that often try to link the origin of dark matter to other outstanding problems in physics and cosmology.  Amongst others these include, for example, weakly interacting massive particles (WIMPs) (e.g., \cite{WIMPs}), axions \cite{axions}, sterile neutrinos \cite{sterile}, SuperWIMPs \cite{SuperWIMPs},  and asymmetric dark matter (ADM) models that link the origin of dark-matter and baryon number (e.g., \cite{ADMBaryon}).  Many of these models are already well constrained from a remarkable effort in experimental dark matter physics including direct and indirect searches for dark matter and constraints from cosmology, and improved measurements and data from the Large Hadron Collidor (LHC) have the potential to further shed light on dark matter physics (see, e.g.,  Ref.~\cite{Feng:2010gw}).

Despite these intriguing possibilities for dark matter physics it remains entirely possible that dark matter belongs to hidden sector and lacks standard model gauge charge.   In such a case dark matter is expected to have at most extra-weak interactions with the standard model, and in the extreme case interacts only via gravitational forces.  In such a case is dark matter physics wholly inaccessible to us?  We take heart in the knowledge that the only reason we know about dark matter is via its gravitational effects in cosmology and astrophysics.  

The most scrutinized candidate for dark matter is arguably the WIMP because the so-called ``WIMP miracle" ensures that the simple process of its weak-scale interactions freezing out from a thermal plasma give it nearly the right relic abundance to be all of the dark matter.   But as this result only relies on the size of the total annihilation cross section being close to  typical electroweak value $\sigma_{\rm EW} \sim 10^{-8}~{\rm GeV}^2$ (and not that the annihilation products are standard model particles) the WIMP miracle can be extended to a ``WIMPless miracle'' in hidden-sector models that also have a similar value for the cross section but with a range of masses extending down to $O(10)$\,keV \cite{Feng:2008mu}.  In Ref.~\cite{Feng:2008mu} a gauge mediated model of supersymmetry breaking is adopted that naturally gives $\sigma \simeq \sigma_{\rm EW}$ and WIMP-like dark matter that decouples from the hidden plasma when it is nonrelativistic. Also see Ref.~\cite{Cheung:2010gj} for different possible origins of hidden sector dark matter. In Ref.~\cite{Sigurdson:2009uz} hidden-sector models that decoupled when ultrarelativistic were considered. In this work we present a unified treatment of the freeze-out of thermal relics in hidden sectors for arbitrary cross section $\sigma$, only requiring that constraints from cosmology are satisfied, and find viable dark matter that freezes out when it is relativistic, semirelativistic, or nonrelativistic.

Already, there are stringent bounds on hidden sector dark matter from cosmology if the hidden-sector reheated to a temperature at or above the temperature of the visible sector.  If the hidden sector has more than a few light species for dark matter to annihilate into then constraints on the number of relativistic degrees of freedom from big bang nucleosynthesis (BBN) rule out such models \cite{Ackerman:2008gi,Simha:2008zj,Izotov:2010ca}.
However, in hidden sector models the reheat temperature of the hidden sector can easily be different from that of the visible (standard model) sector depending on the details of the reheating process or if the visible sector and hidden sector contain a different particle spectrum (different number of degrees of freedom as a function of temperature) and thus cool differently.   Note that implicit in saying the visible and hidden sectors are at different temperatures is the assumption that the two sectors  are not in thermal contact (or at least they lost thermal contact considerably before BBN) and processes that could thermalize them in a Hubble time are inefficient. 

As discussed above, we take a model-independent approach to hidden-sector dark matter in this paper and let the cross section $\sigma$ be a free parameter.  In addition we allow the mass $m_{\chi}$ of our dark matter particle $\chi$ and the hidden-to-visible temperature ratio $\xi=T^h_f/T_f$ (at dark matter freeze-out) be additional parameters and map of the viable region of this three-dimensional parameter space.  Depending on the values of $m_{\chi}$, $\xi$, and $\sigma$ the 
 freeze-out process can be relativistic, semirelativistic or nonrelativistic.  
 
Our paper is organized as follows:  In Sec.~\ref{sec:ra} we discuss the standard Boltzmann equations involved in the freeze-out process and how they must be extended for the general hidden-sector case with $\xi < 1$.  In Sec.~\ref{sec:ta} we discuss our approach to modelling the temperature dependence of the annihilation cross section in detail, how we treat the problematic semirelativistic regime, and a numerical approach to relativistic and nonrelativistic decoupling within the same framework.   In Sec.~\ref{sec:fs} we calculate bounds arising from cosmological limits on the free-streaming length of dark matter, while in Sec.~\ref{sec:tg} we calculate limits from the
 generalized Tremaine-Gunn bound on the phase space density of dark matter.  
 We discuss our results in Sec.~\ref{sec:results}, the strength of hidden-visible interactions in Sec.~\ref{sec:detection}, and conclude in Sec.~\ref{sec:conclusion}.
 
\section{ The Relic Abundance}
\label{sec:ra}
In this section  we review the standard Boltzmann equation for the freeze-out of annihilation of dark matter and its extension to the hidden-sector case.  Suppose we have dark-matter particle of mass $m_{\chi}$ interacting with a plasma of other particles at temperature $T$. In terms of the yield, or number density to entropy ratio, $Y= {n_{\chi}}/{s}$ we can write
\begin{equation}
\label{basicboltz}
\frac{dY}{dx}=- \frac{1}{x^2} \frac{s(m_{\chi})}{H(m_{\chi})} \sigv \left[Y^2 - Y_0(x)^2\right] \, ,
\end{equation} 
where $x\equiv m_{\chi}/T$ a convenient time parameter, $Y_0(x)$ is the equilibrium yield (the yield if dark-matter annihilation were in equilibrium), and $s(m_{\chi})$ and $H(m_{\chi})$ are the total entropy and Hubble rate evaluated at $T=m_{\chi}$ respectively.\footnote{We use units with $c=1$ and $k_{B}=1$} In the Maxwell-Boltzmann (MB) approximation we can write \cite{Gondolo:1990dk}

\begin{equation}
Y_0(x) = \frac{45}{4 \pi^4} \, \frac{d_{\chi}}{g^{s}(x)} x^2 K_2(x) \,
\end{equation}
where $d_{\chi}$ is the number of internal  freedoms of a dark-matter particle, $g^{s}(x)$ is the effective number of entropy degrees of freedom, and here and elsewhere, $K_{\rm n}(x)$ are modified Bessel functions. 
This can be expressed in a convenient form by using the scaled variable
$y\equiv[{s(m_{\chi})}/{H(m_{\chi})}] \langle\sigma v\rangle Y$.  In the limit that the temperature dependence of $\sigv$ is negligible we then have 
\begin{equation}
\frac{dy}{dx}=-\frac{1}{x^2}\left[y^2 - y_0^2(x)\right]
\label{maineq}
\end{equation}
where $y_0$ is the equivalently-scaled equilibrium yield.  If dark matter is part of the visible sector and decouples when nonrelativistic then standard freeze-out results are found by solving Eq.~(\ref{maineq}) with $y_0(x)$ as the initial condition for $y$ at small $x$ (high temperatures).

However, if dark matter is part of a hidden sector decoupled from the visible sector when it freezes out then its equilibrium distribution is given by the hidden-sector temperature $T_{h}=\xi T$ rather than $T$, where we have introduced the hidden-to-visible temperature ratio $\xi$.
It's conventional to cast the total density (that appears in the Hubble rate) or entropy density in terms of an effective number of degrees of freedom $g(T)$ [or entropy degrees of freedom $g^{s}(T)$] at a temperature $T$.  As discussed in, for example, Refs.\cite{Feng:2008mu,Sigurdson:2009uz}, in a universe with both a visible and hidden sector there are two copies of these functions, $g_{v}(T)$ for the visible sector and $g_{h}(T_{h})$ for the hidden sector.  Since for relativistic particles the energy density scales as $\rho \propto T^4$ and entropy density scales as $s \propto T^3$ we find that the combined $g(T)$ and $g^{s}(T)$ take the form
\begin{align}
\notag g(T) &\equiv g_v(T) + g_h(\xi T)\xi^4 \simeq g_v(T) \\ 
g^{s}(T) &\equiv g^{s}_v(T) + g^{s}_h(\xi T)\xi^3 \simeq g^{s}_v(T) \, ,
\end{align}
where the rightmost approximation typically holds when $\xi \lesssim 0.5$ and is used it in the remainder of the paper.  If the hidden sector is relatively cold compared to the visible sector then the energy density (and Hubble expansion rate) of the Universe is essentially given by the density of visible-sector radiation.  Thus the key change to Eq.~(\ref{basicboltz}) is just that $Y_0(x) \rightarrow {Y}_0(x;\xi)$, where
\begin{equation}
{Y}_0(x;\xi) \equiv \xi \frac{45}{4 \pi^4} \, \frac{d_{\chi}}{g^{s}(x)} x^2 K_2(x/\xi) \,
\end{equation}
is the equilibrium yield of dark matter in a hidden sector and the standard case is recovered when $\xi \rightarrow 1$.  In terms of the scaled yield $y$ we have
\begin{equation}
\frac{dy}{dx}=-\frac{1}{x^2}\left[y^2 - y_0^2(x;\xi)\right] \,
\label{eq:constcross}
\end{equation}
with $y_0(x;\xi) \equiv[{s(m_{\chi})}/{H(m_{\chi})}] \langle\sigma v\rangle Y_0(x;\xi)$.
Equation (\ref{eq:constcross}) is valid when the thermally averaged cross section is almost constant at freeze-out.
We discuss in the next section the more general case when the temperature dependence of the cross section cannot be neglected.

\section{ Thermally averaged cross section }
\label{sec:ta}
While approximate analytical solution are well known when dark matter decouples when either highly relativistic $ x_f \equiv m_{\chi}/T_f \ll 1 $ or nonrelativistic $x_f \geq 3$,
for the intermediate semirelativistic case the thermally-averaged cross section $\sigv$ cannot be expanded in either dark matter mass or velocity, and analytical methods are less tractable.
However, in Ref.~\cite{Drees:2009bi}, a useful and accurate interpolating expression for the thermal average cross section was introduced that can be used to track the dark matter freeze-out process even
in the semirelativistic regime $x_f \simeq 1$. 
 The general expression for $\sigv$ is given by 
 \begin{equation}
\sigv = \frac{1}{8 m_{\chi} T K_2^2(x)} \int_{4 m_{\chi}^2}^{\infty} ds \,\sigma (s - 4 m_{\chi}^2 )\sqrt{s} K_1(\sqrt{s}/T)
\end{equation} 
While, strictly speaking, using the MB distribution is improper for the semirelativistic case, it has been shown to yield accurate results for the final relic density \cite{Drees:2009bi}.
This is partly due to cancellations between the numerator and denominator and also because the final relic density becomes quite insensitive to 
the precise value of $x_f$ as freeze-out moves from the nonrelativistic toward the relativistic case. 
In the case of an s-wave process,\footnote{The p-wave case can be treated similarly although the approximating ansatz is different \cite{Drees:2009bi}.} after some simplification, $\sigv$ can be put in the form 
\begin{equation}
\sigv= \sigma  \frac{4}{x^6 K_2^2(x)} \int_0^{\infty} dt \, \,  t^2 \, (t^2 + x^2)^2 K_1(2\sqrt{t^2 + x^2})
\end{equation}
Now in the nonrelativistic and ultrarelativistic limits it reads 
\begin{align}
\sigv_{NR} \equiv \lim_{x \gg 3} \sigv = \sigma  \notag \, \,\,\,\,\,\, {\rm and}&\, \\
\sigv_{R}  \equiv \lim_{x\ll 1} \sigv = \frac{\sigma}{4 x^2}(12 + 5 x^2)&\,  
\end{align}
repectively. A simple ansatz that interpolates between the these cases is \cite{Drees:2009bi}
\begin{equation}
\sigv \equiv \sigma f(x) \equiv \sigma \left(\frac{3}{x^2} + \frac{\frac{5}{4}+x }{1+x}\right) \, ,
\end{equation}
where $f(x)$ is chosen to smoothly reproduce both the  ultrarelatvistic ($x \ll 1$) and nonrelativistic ($x \gg 3$) cases.  Incorporating this new temperature-dependent cross section $\sigv = \sigma f(x)$ 
we find Eq.~(\ref{eq:constcross}) takes the form 

\begin{equation}
\frac{dy}{dx}=-\frac{1}{x^2}\left[y^2 - y_0^2(x;\xi)\right] + y \, \frac{\, \,d\ln{f(x)}}{dx} \, .
\label{eq:fulleq}
\end{equation}
\begin{figure}[t]
\centering
\includegraphics[scale=0.8]{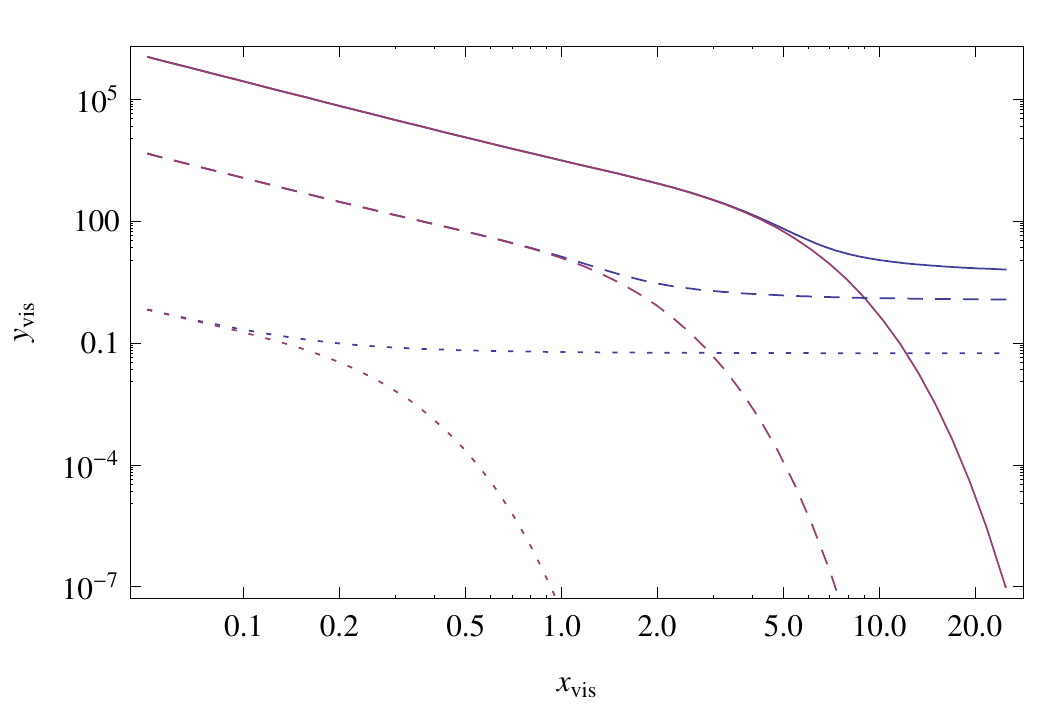}
\caption{The (scaled) yield $y(x)$ and equilibrium yield $y(x;\xi)$, where $x_{vis}=x=m_{\chi}/T$ represents visible sector temperature,  are shown for different choices of $\sigma$ and $\xi$.   The solid, dotted and dashed curves represent 
$\xi=0.9$, $\xi=0.3$, and $\xi=0.05$ respectively for $\sigma= 0.1 \sigma_0$ and $\sigma_0 = 10^{-9}\,\,{\rm GeV}^{-2}$.
In agreement with Ref.~\cite{Feng:2008mu}, increasing $\xi$ at fixed $\sigma$ increases the relic density. }
\label{xf}
\end{figure} 
With the above approximation, we can  numerically solve Eq.~(\ref{eq:fulleq}) for the freeze-out process as a function of $\sigma$, $m_{\chi}$, and $\xi$. A key quantity we need to know to estimate the final relic abundance and whether it occurs in the ultrarelativistic, semirelativistic, or nonrelativistic regimes is the freeze-out temperature $T_f$ or equivalently $x_f= {m_{\chi}}/{T_f} $.   Exactly how $T_f$ is defined is somewhat arbitrary, but a natural definition is the temperature $T_f$ at which $\Gamma < H$ for the first time --- the interaction rate drops below the Hubble rate.   Such a definition can be used to accurately estimate the final yield in the relativistic and semirelativistic cases, and can naturally extend to the conventional nonrelativistic case.  Explicitly, we use the condition
\begin{equation}
\left[s Y \sigv\right]_{T_f} = H|_{T_f} \, ,
\end{equation} 
which in terms of the scaled yield becomes simply $y(x_f)=x_f$.  We use this later condition to determine $x_f$ numerically.  
The freeze-out solution is shown in Fig.~\ref{xf} for 3 different choices of $\xi$ at fixed $\sigma$, where we see higher $\xi$ results in higher relic density. Despite higher values of $x_f$, and thus later freeze-out with more exponential suppression, the larger relative temperature more than compensates in this example. 

Although reasonable approximations can be found in the nonrelativistic and ultrarelativistic cases, to calculate the present relic density in what follows we use the yield found from numerically solving Eq.~(\ref{eq:fulleq}) at a late time $x_{\infty}$
where the yield is essentially constant.  The cosmological dark-matter density is then
\begin{equation}
\Omega_{\chi}= m_{\chi} s_0 \frac{Y(x_{\infty},\xi)}{\rho_c} \,
\end{equation}
where $\rho_c$ is the critical density of the Universe.
We use the measured value of the dark matter relic density from Ref.~\cite{Dunkley:2008ie} to constrain the parameter space.

\section{Bound from Free-streaming}
\label{sec:fs}
Generally, hidden-sector dark matter does not need to freeze-out when nonrelativistic (cold) in order to match observations \cite{Sigurdson:2009uz}.   The dark-matter particles may have significant thermal momentum at freeze-out (it can be warm or hot from the hidden-sector perspective), and this can lead to a significant free-streaming length that can suppress the matter power spectrum on scales as large as those corresponding to galactic length scales.  While the freeze-out process is thermal in the hidden sector, since the hidden sector is at a different temperature than the visible sector, it mimics a non-thermal freeze-out process.   While the detailed particle physics model can be rather different, it is quite similar to the familiar case of sterile neutrino warm dark matter \cite{Boyarsky:2008mt,Abazajian:2006yn, deVega:2009ku,deVega:2011gg,deVega:2011gs,Boyanovsky:2007ay,Kusenko:2010ik,Ando:2010ye,deVega:2011xh} with the important difference that it did in fact thermalize in the hidden sector at a temperature lower than the visible universe. We will see that this difference will make the free-streaming bound somewhat less stringent compared to the a standard warm dark matter case with $\xi=1$.

\begin{figure}[t]
\includegraphics[scale=0.8]{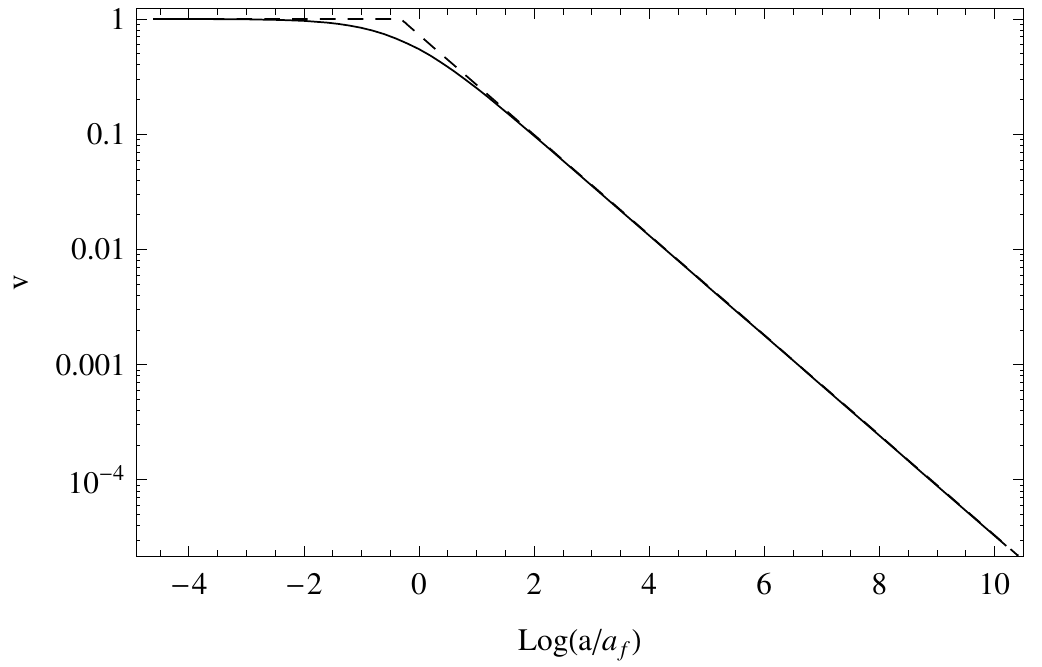}
\caption{ The solid line shows the actual dependence of particle velocity as a function of $\bar{a} = {a}/{a_f}$ for  $(m_{\chi}=2.5 \rm{keV}, x_f=2, \xi= 0.3)$ while the dashed line shows the approximation we have used for our numerical code.}
\label{Contours12345}
\end{figure}

\begin{figure*}[t]
\centering
\includegraphics[width=2.2in]{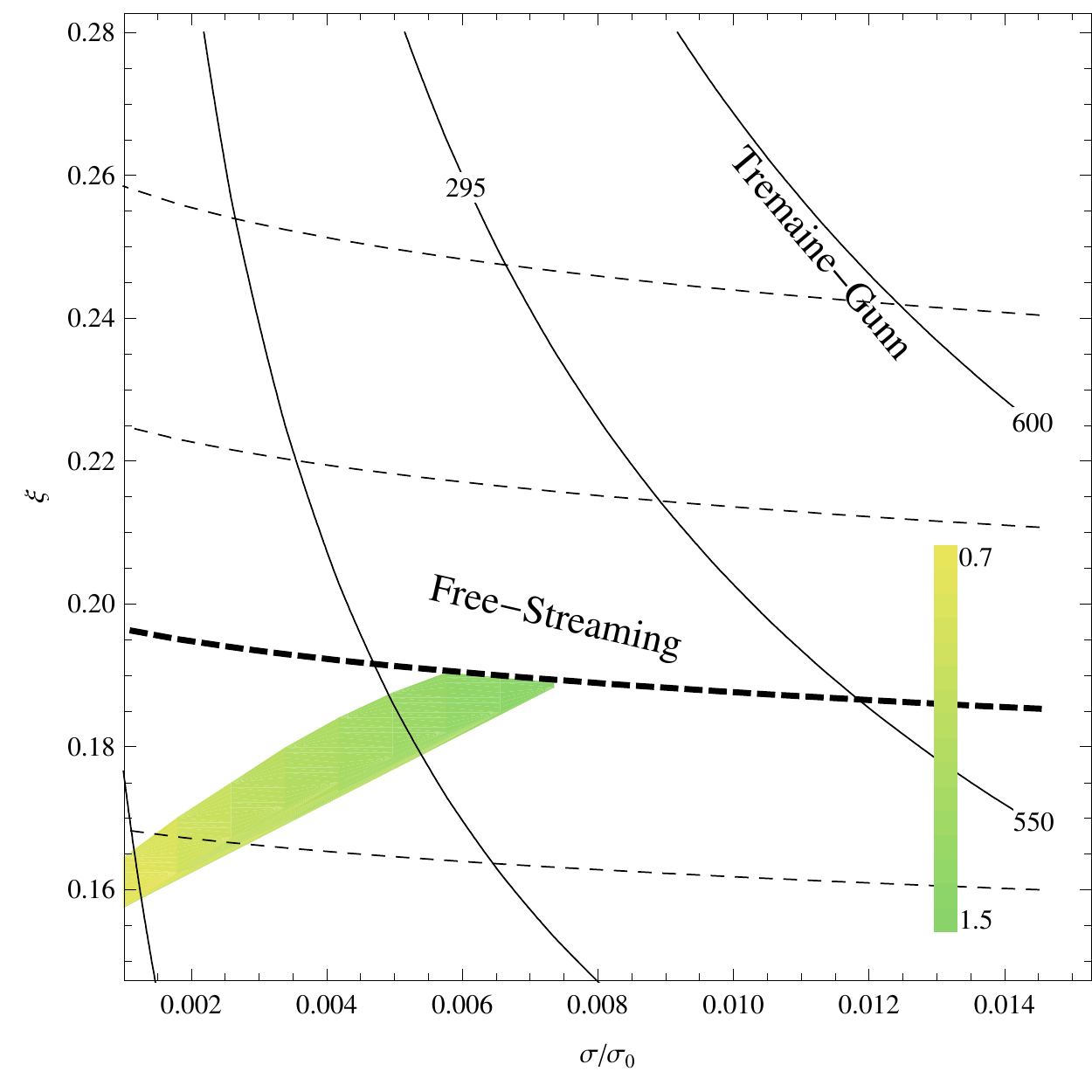}\quad
\includegraphics[width=2.2in]{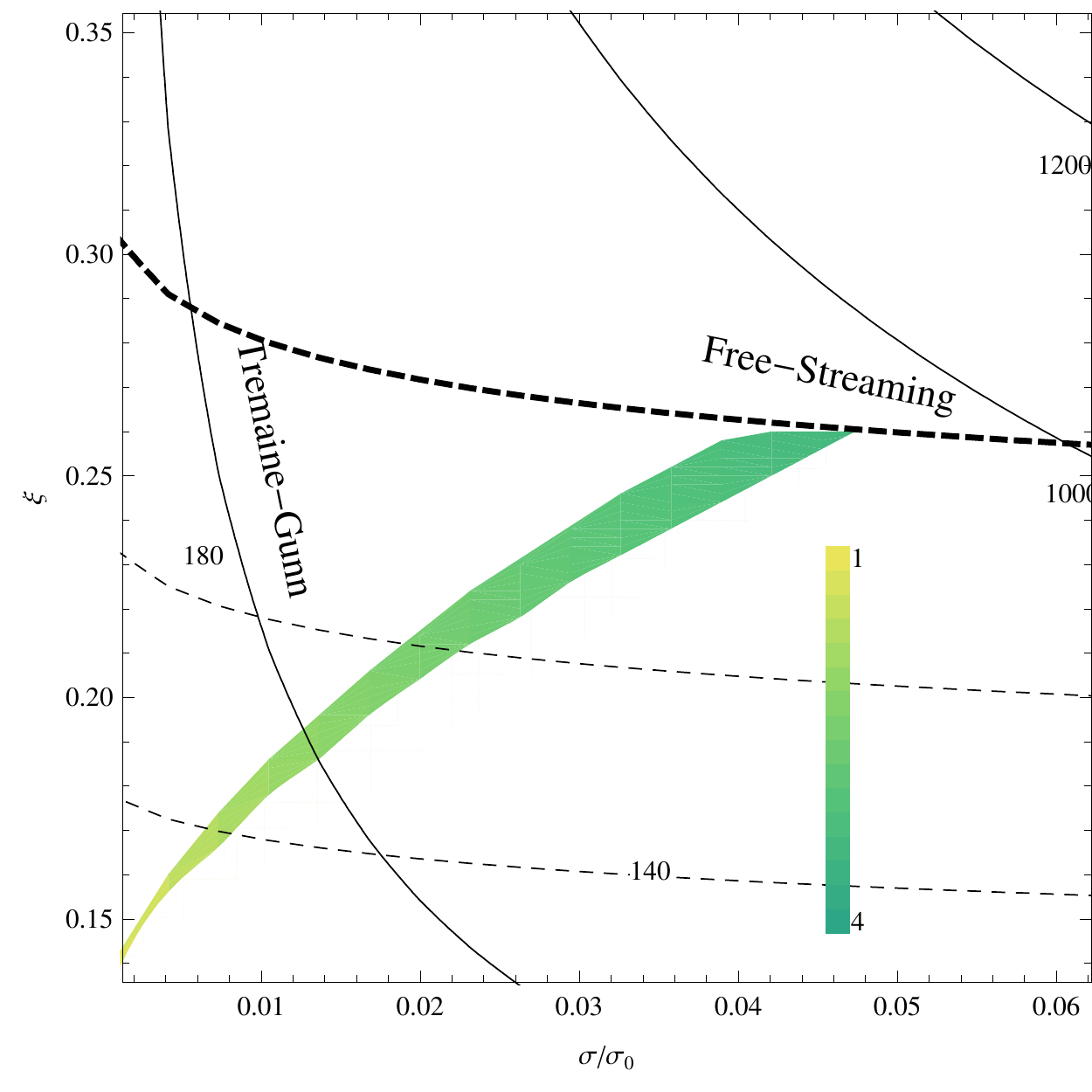}\quad
\includegraphics[width=2.2in]{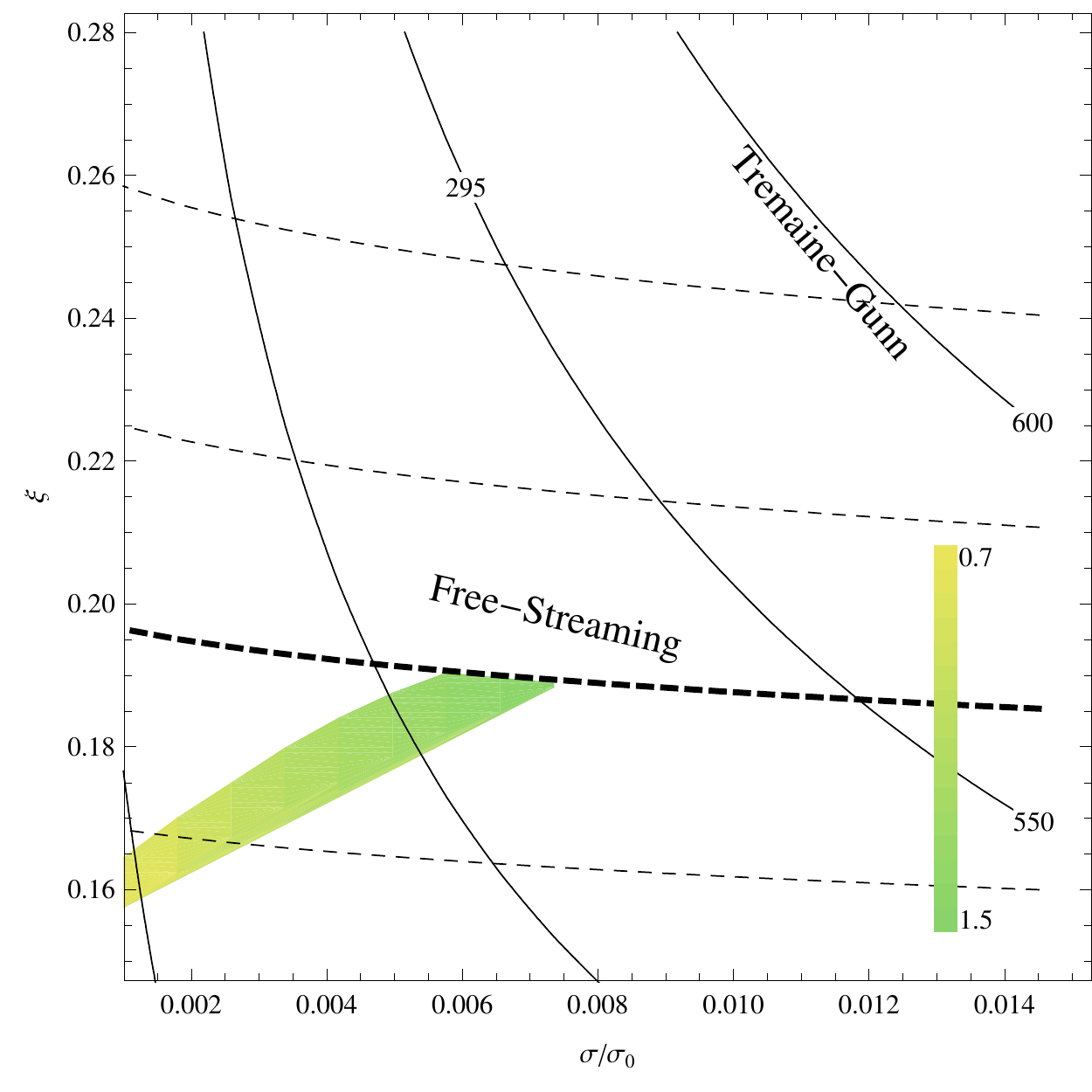}\quad
\caption{Representative constraints in $({\sigma}/{\sigma_0}, \xi)$ plane for \mbox{$m_{\chi}=1900$\,\,{\rm keV}} (left panel), \mbox{3\,\,\rm{keV}} (middle panel), and \mbox{$12.5$\,\,\rm{keV}} (right panel).  Here, $\sigma_0=10^{-9}$ $\rm GeV^{-2}$. The horizontal dashed
 lines are contours of free-streaming length while the bold one corresponds to  $ \lambda^{FSH} \simeq 230\,\,\rm{kpc}$. The solid lines are contours of minimum Tremaine-Gunn mass, $m_{min}^{TG}(\xi,\sigma)$, and the bold solid line (right panel) corresponds to $m_{min}^{TG}( \xi,\sigma )= m_{\chi}$. We see that the free-streaming bound is more stringent for low dark matter mass, while the Tremaine-Gunn  bound restricts the parameter space for higher $m_{\chi}$ (e.g., right panel).  The colored region corresponds to the allowed values of dark matter density ($ 0.1\leq \Omega_{DM} h^2 \leq 0.114$) and that also obey the Free-streaming and Tremaine-Gunn bounds. The corresponding values of hidden-sector 
$x_f^h \equiv m/T_f^h$ are shown in the sidebar to illustrate the nature of decoupling (i.e., whether relativistic, nonrelativistic or semirelativistic). We see that for low mass $x_f^h $ is smaller and decoupling tends to be relativistic, while for higher mass $x_f^h $ is larger with values corresponding to either semirelativistic (warm) or nonrelativistic freeze out. }\label{Contours123}
\end{figure*}

It was pointed out in Ref.~\cite{Boyarsky:2008xj} that in the case of warm dark matter, a common approximation for the free-streaming scale used for standard model neutrinos, \,\,
$ \lambda_{fs}(t) = {2 \pi}/{k_{fs}(t)}$, does not correspond to the actual suppression scale in the matter power spectra. 
Instead, the largest scale affected due to free-streaming is essentially the particle horizon of dark matter particles.  The comoving free-streaming horizon of a dark matter particle with typical velocity $\ave{v}$ is given by
\begin{equation}
\lambda^{FSH}= \int_{t_f}^{t_{eq}} \frac{\ave{v}}{a} dt 
\end{equation}
where $t_f$ and $t_{eq}$ corresponds to time of freeze-out and matter-radiation equality respectively and $a$ is the scale factor. 
The scale factor dependence of the velocity changes with  $x_f$ and the mass of dark matter.  A simple calculation shows the momentum average of the particle velocity,
 $  \ave{v} = \ave{ {p}/{\sqrt{m^2 + p^2}}}$, is given by 

\begin{equation}
\ave{v}=\ave{v(\bar{a})} \equiv \frac{\int_0^{\infty} dq \, q^2 \left(\frac{q}{\sqrt{q^2 + \bar{a}^2 m^2}}\right) f(q,T^h_f) } {\int_0^{\infty} dq \, q^2 f(q,T^h_f)}
\label{vel}
\end{equation}
$q \equiv p_f = \bar{a}p $ is the momentum at freeze-out, and $T^h_f=\xi T_f$ is the hidden-sector temperature at freeze-out.  Here $p$ is the physical momentum, $\bar{a} ={a}/{a_f}$, $a_f$ is the scale factor at freeze-out,  and \mbox{$f(q,T^h_f)=[1+ \exp(\sqrt{m^2+q^2}/T^h_f)]^{-1}$} is the Fermi-Dirac distribution at the hidden-sector freeze-out temperature.
At early times $\bar{a} \rightarrow 0$ and, provided the freeze-out temperature is high enough, $ \ave{v} \rightarrow 1$  as expected. But as $\bar{a}$ increases and the momentum of each particle decreases we find $\ave{v} \propto \bar{a}^{-1}$.
 
While Eq.~(\ref{vel}) quantifies exactly how to calculate the thermal average, a simple approximation can be used to calculating $\lambda^{FSH}$ to sufficient accuracy.
We split the free-streaming epoch into two regimes by defining a transition scale factor $a_{nr}$ when dark matter particles become nonrelativistic after freeze-out at $a_f$, and write $\lambda^{FSH}$ as a sum of two integrals
\begin{equation}
\lambda^{FSH}= \frac{1}{\sqrt{\Omega_r} H_0}  \left[\int_{a_{f}}^{a_{\lambda}} da  +  \int_{a_{\lambda}}^{a_{e}}\frac{ a_{nr} \, da}{\sqrt{a^2+ \frac{a^3}{ {a_{e}}}}} \right] \, .
\end{equation}
Within our approximation, prior to $a_{nr}$ we take $\ave{v} = 1$ while for $a > a_{nr}$ we take $\ave{v} \propto {a}^{-1}$, and $a_{\lambda} = \max[a_{f},a_{nr}]$. Here, $a_{nr}$ corresponds to a scale factor when dark matter particles becomes nonrelativistic ($T^h_{nr} \simeq {m_{\chi}}/{3.15} $) \cite{Sigurdson:2009uz}. This corresponds to a visible sector temperature $T_{nr} \simeq ({m_{\chi}}/({3.15 \xi})) ({g^s_f}/{g^s_{nr}})^{{1}/{3}}$.  For particles that freeze out when already nonrelativistic, so that $a_f >  a_{nr}$, only the second term contributes.  In Figure~\ref{Contours12345}, we plot the numerical solution of free-streaming velocity as a function of scale factor and also show the approximation we have adopted. We find the free-streaming length computed using this approximation versus Eq.~(\ref{vel})  agree at the few percent level --- accurate enough for this work.

For relativistic decoupling it is straightforward to show that the free-streaming length is only a function of mass, but for the semirelativistic and nonrelativistic cases it depends on both mass and $\xi$.  
In all cases the comoving free-streaming length must satisfy the bound from the measurements of the linear power spectra on small scales which in turn constrains the dark-matter parameter space  $(\xi, \sigma, m_{\chi})$. 
We use here the bound \mbox{$\lambda^{FSH} \leq 230~{\rm kpc}$}  \cite{Boyarsky:2008xj,Sigurdson:2009uz}. 

\section{Tremaine-Gunn Bound}
\label{sec:tg}
A robust and model independent lower bound on the mass of dark matter particles is obtained by bounding the phase-space density evolution of small galaxies like the dwarf spheroidal satellites (dSphs) of the Milky Way. Originally suggested by Tremaine and Gunn in Ref.~\cite{Tremaine:1979we}, it has been developed further by several authors in, for example, Refs.~\cite{Madsen:1983vg,Madsen:1991mz,Madsen:1990pe,Boyarsky:2008ju}. If the dark matter is fermion it gets a stringent bound, independent of cosmological evolution, as Pauli blocking enforces a densest packing of the dark matter phase space distribution.  However, a different bound relevant for both bosons and fermions applies to any bath of particles once in thermal equilibrium with interactions that freeze out.  Since the microscopic phase space density (PSD) is exactly conserved for collisionless and dissapationless particles by Liouville's theorem the coarse-grained (averaged) PSD can not be an increasing function of time.  This means that the maximum coarse-grained PSD in a galaxy today, with core radius $r_c$ and velocity dispersion $\sigma_{\rm v}$, must not exceed the maximum value of the microscopic PSD (the Fermi-Dirac  for a fermion) over cosmic history $f^{cg}_{max}(t) \leq f^{fg}_{max}$.  
This inequality translates to a bound on the mass of

\begin{equation}
m \geq m_{min} = \left(\frac{9 \, h^3}{ (2 \pi)^{5/2} d_{\chi} \,  G_N \, \sigma_{\rm v} \, r_c^2 } \frac{1}{ f(q,T^h_f)|_{max}} \right)^{1/4}
\end{equation} 
where $f(q,T^h_f)$ is defined in the previous section. Here one assumes that the collapse of dark matter from the initial state to bound halos is dissipationless and collisionless. The maximum of  of the distribution $f(q,T^h_f)$  is a function of $x_f^h$, and through this the Tremain-Gunn mass $m_{min}$ also depends on $\xi, m_{\chi} $ and $\sigma$.  There has been many observational studies employing the Tremaine-Gunn lower bound on dark matter mass. As a representative example we use the results of the ultra-faint dwarf spheroidal galaxy Leo \rm{IV}  \cite{Simon:2007dq, Martin:2008wj} to restrict our parameter space.

\begin{figure*}[t]
\centering
\includegraphics[width=2.2in]{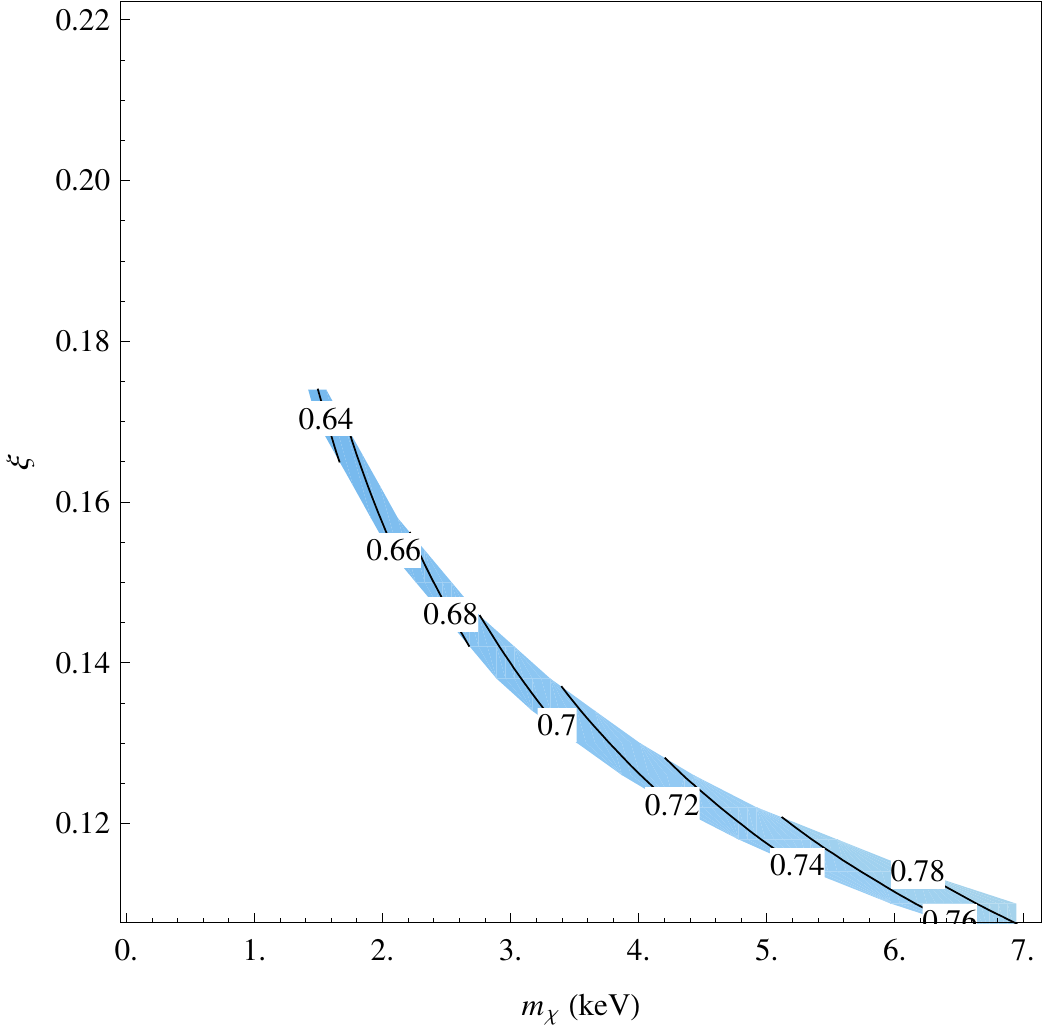}\quad
\includegraphics[width=2.2in]{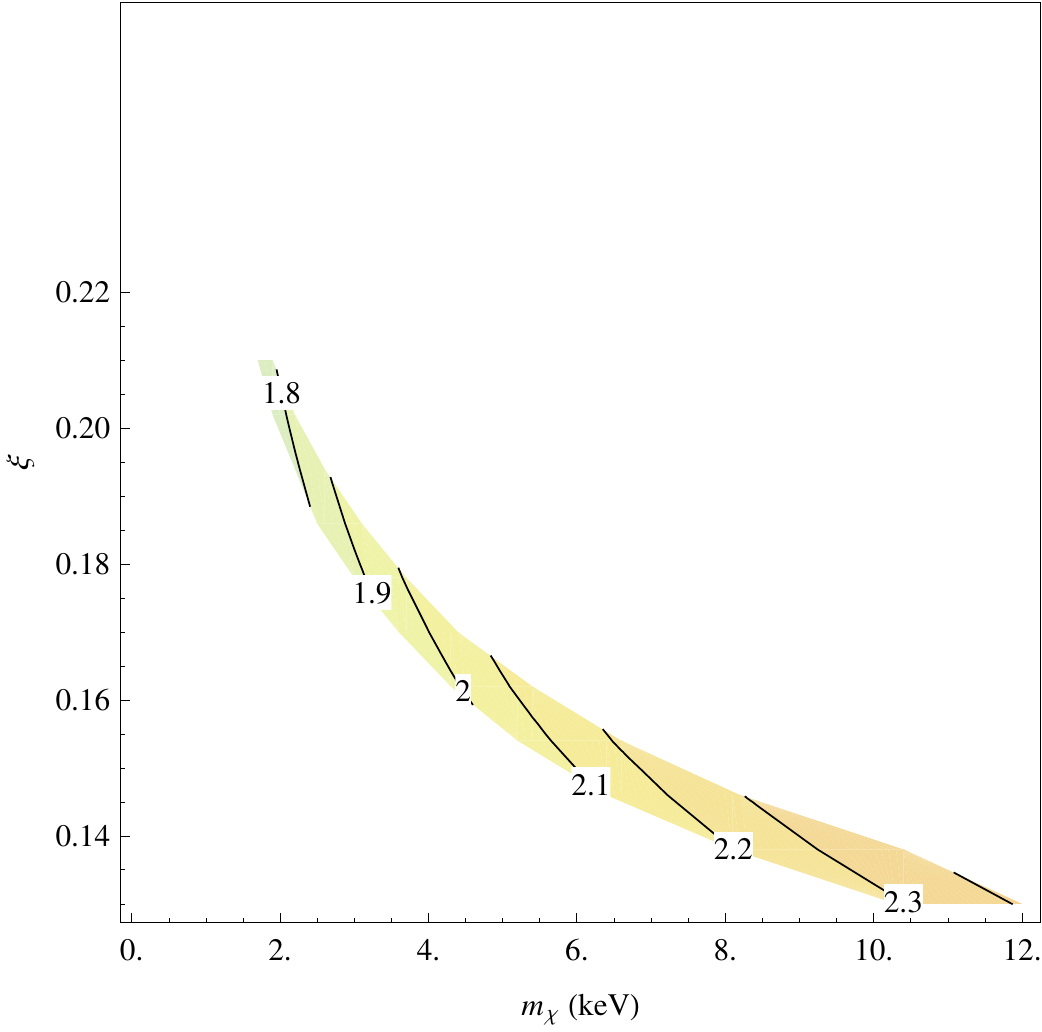}\quad
\includegraphics[width=2.2in]{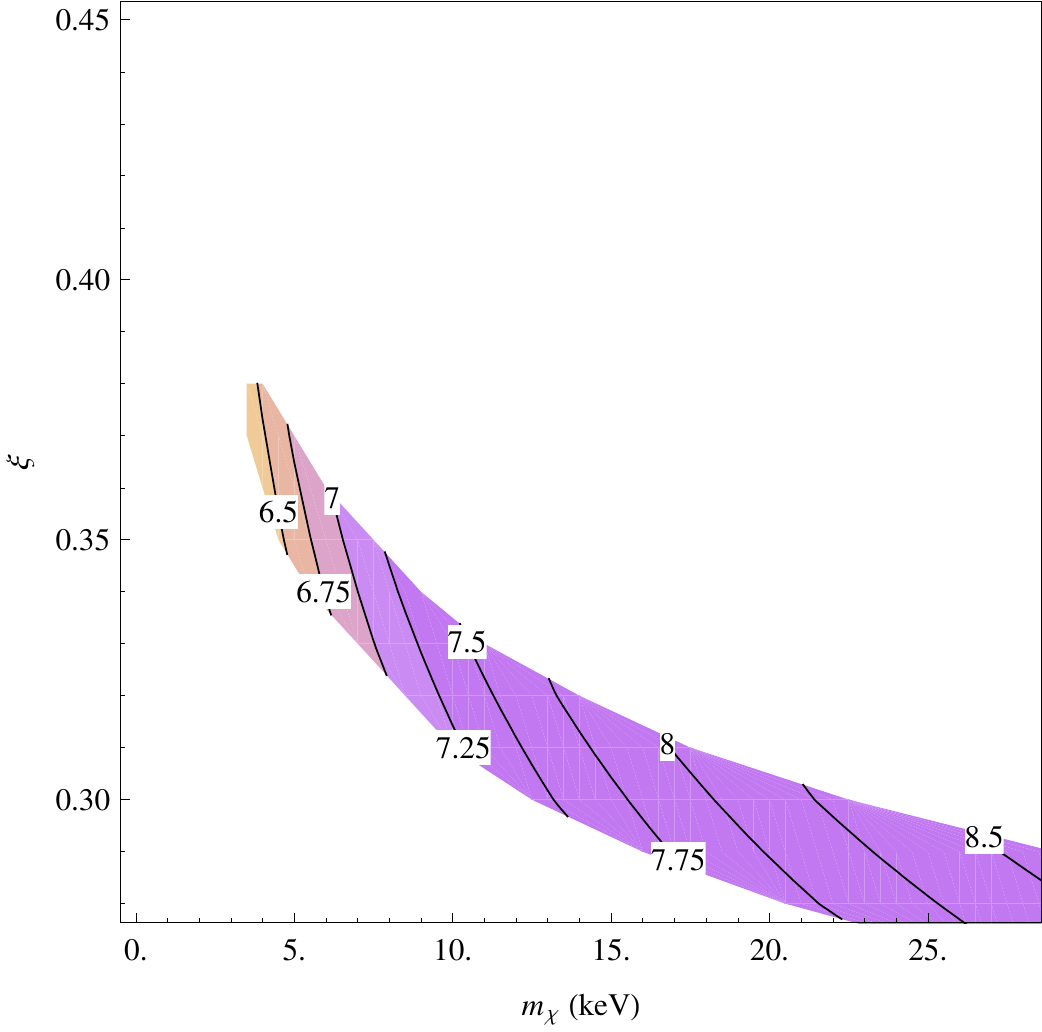}
\caption{Representative constraints in the $(m_{\chi}, \xi)$ plane for ${\sigma}/{\sigma_0}= 10^{-3}, 10^{-2}$ and $10^{-1}$. Where $\sigma_0=10^{-9}$ ${\rm GeV}^{-2}$. The colored region corresponds to  the allowed values of  dark matter density ($ 0.1\leq \Omega_{DM} h^2 \leq 0.114$) that also obey free-streaming and Tremaine-Gunn bounds. The corresponding values of the hidden-sector 
$x_f^h \equiv m/T_f^h$ has been shown to illustrate the nature of decoupling (i.e., whether relativistic, nonrelativistic or semirelativistic). We see for smaller mass freeze-out tends to be relativistic and for larger mass it is semirelativistic or nonrelativistic. }\label{newContours}
\label{fig:fixedmass}

\end{figure*}

\begin{figure}[b]
\includegraphics[width=2.32in]{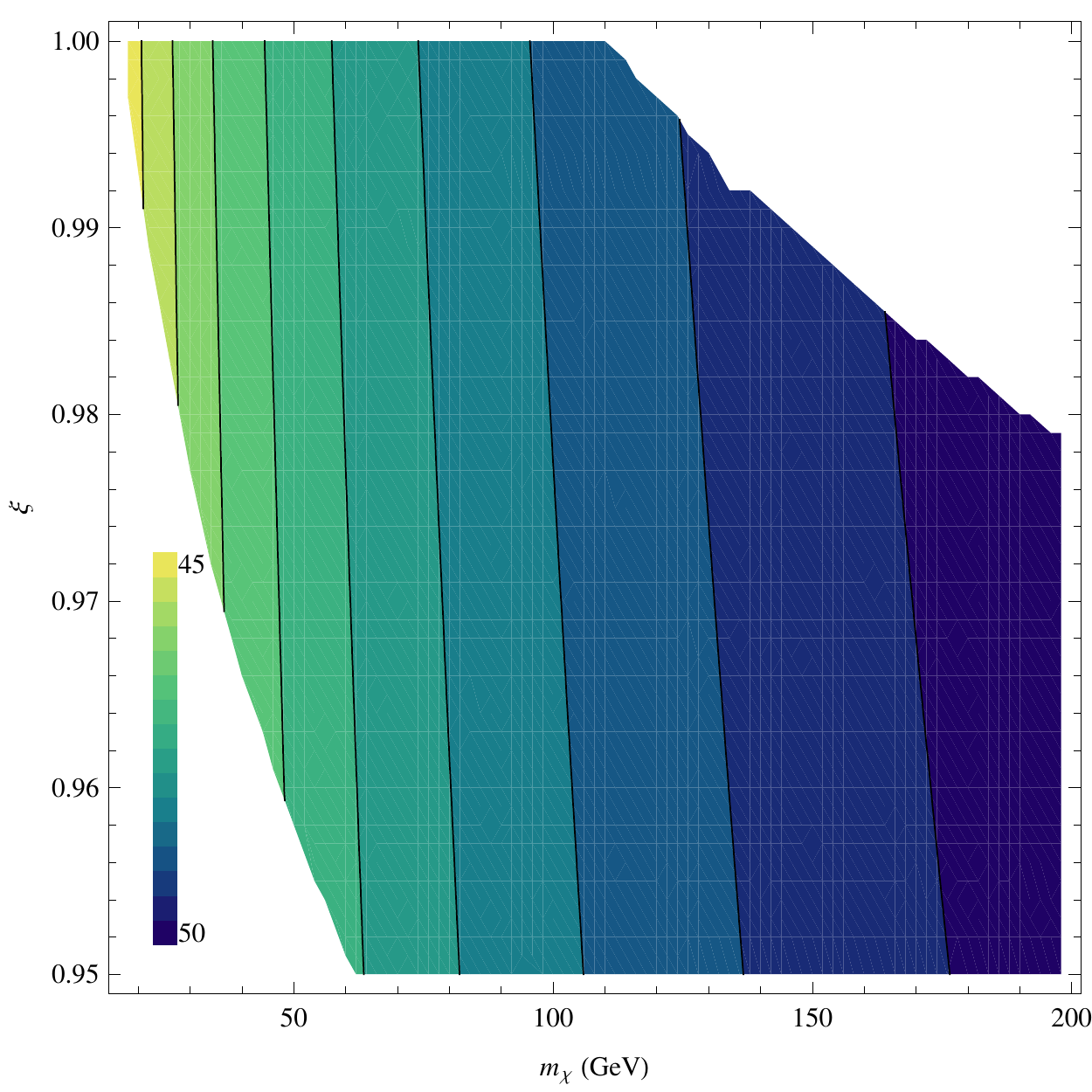}
\caption{ The allowed region in the $(m_{\chi}, \xi)$ plane for \mbox{$\sigma= 1.5 \times 10^{-8}~ {\rm GeV}^{-2}$}. As this cross section is of the same order of a canonical cold WIMP, the allowed region extends all the way up to $\xi=1$ as expected for $ m_{\chi} \sim 100~{\rm GeV}$.}\label{Contourswimp}
\end{figure}

\section{Results}
\label{sec:results}	
We now discuss our results for the allowed parameter space in the rather general hidden dark matter models discussed in this work.  We performed a survey of the parameter space in the ($\xi,\sigma$) plane for fixed $m_{\chi}$ and in the ($\xi,m_{\chi}$) plane for fixed $\sigma$ --- solving the freeze-out equations numerically.   We then require that hidden-sector dark matter: i) accounts for the observed dark-matter density; ii) has a free-streaming length consistent with structure formation; and iii)  satisfies the Tremaine-Gunn PSD bound. 
Our results extends the work of Ref.~\cite{Feng:2008mu} for several reasons. Firstly, we allow the size of the dark matter scattering cross section to vary.  Secondly, freeze-out and decoupling does not necessarily occur when the dark matter particles are nonrelativistic $(x_f^h \geq 3)$.   Finally, since these constraints are relevant for relativistic and semirelativistic models, we include constraints from the free-streaming length and PSD.  A special case of our study with $\sigma = \sigma_{EW}\sim 90\sigma_0$ and $x_f^h \geq 3 $ corresponds to WIMPless dark matter scenario studied in Ref.~\cite{Feng:2008mu}, and we have checked that our calculation reproduces their results in that limit.

In Fig.~3 we show the allowed parameter space in the ($\xi,\sigma$) plane for different choices of dark-matter mass. We find that for dark matter masses $m_{\chi} \leq 1.5 \, \rm{keV}$ there are no values of $\xi$ or $\sigma$ that satisfy all the three constraints. This puts a lower bound $m_{\chi} \gtrsim 1.5~{\rm keV}$ on the mass of hidden-sector dark matter that can account for all  dark-matter in the Universe in a model-independent way.  This bound is weaker than the limit on conventional ($\xi=1$) wark darm matter (WDM) of $m \gtrsim 2~{\rm keV}$ described in Refs.~\cite{Seljak:2006qw,Viel:2006kd,Boyarsky:2008mt} because the hidden sector is at a lower temperature ($\xi < 1$).

 It is clear from Fig.~3 that at very low mass the free-streaming bound is more important than the Tremaine-Gunn bound, and so the lower limit on mass $m_{\chi} \geq 1.5\,~\rm{keV}$ is determined mainly by the free-streaming bound.  At higher masses the Tremaine-Gunn bound becomes relevant as free-streaming becomes more and more suppressed (see the right panel of Fig.~3) and the allowed region of parameter space  increases. Eventually, for $m_{\chi} \gtrsim 25~{\rm keV}$, only the constraint from dark matter relic density remains relevant. The bands shown represents the width arising from the range of allowed dark matter relic density $ 0.1\leq \Omega_{d} h^2 \leq 0.114$. 
The legend color bar represents different values of ${m_{\chi}}/{T_f^h}$ at the time of freeze-out and shows the transition from relativistic, to semirelativistic, to nonrelativistic freeze-out as $m_{\chi}$ increases.
We see that freeze-out can be easily semirelativistic $1 \lesssim x_f \lesssim 3$ or relativistic $x_f \lesssim 1$ and thus showing hidden-sector dark matter that freezes out when warm or hot is consistent with all cosmological bounds and may be considered when building particle-physics models of dark matter.
    
In Fig.~\ref{newContours} we show the allowed parameter space  in ($\xi,m_{\chi}$) plane for three choices of the scattering cross section $\sigma$.  We see that, intuitively, for lower cross sections and interaction rates the freeze-out process tends to be relativistic, but as the cross section increases freeze-out can occur later and be semirelativistic or nonrelativistic. In Fig.~\ref{Contourswimp} we show that we recover the standard cold WIMP scenario  for higher mass ($\sim\!\!100~{\rm GeV}$) and  typical wimp scale cross section. As we move to higher mass, $\xi$ need to be below unity to satisfy the dark matter abundance.

\section{Nongravitational detection of hidden dark matter?} 
\label{sec:detection}
If the dark matter is in a hidden sector then it has at most feeble interactions with the standard model so that it never equilibrated with the standard model plasma (or at least fell out of equilibrium very early on).  Its direct detection is extremely unlikely and one would expect its detection only through astrophysical signatures.  However, a more precise statement is that any possible interaction rate with the standard model must be less than the Hubble rate $\Gamma_{I}(T) \leq H(T)$ throughout cosmic history. In spite of this constraint, it is not impossible to get direct or indirect detection of hidden-sector dark matter.  This occurs because of extra freedom in the hidden sector freeze-out process.  In contrast with typical WIMP models, a wide range of dark matter mass can produce the observed thermal relic density, and importantly the dark matter annihilation cross section $\sigma$ determines only the thermal relic density while the direct or indirect detection rates are determined by $ \sigma_{I}$. In the context of the WIMPless scenario Ref.~\cite{Feng:2008dz} has suggested that scalar hidden dark matter may have non-gauge interaction with the standard model through connector particles, with interaction of the form  $ {\cal L_I} = \lambda_f X \bar{Y_L} \, f_L + \lambda_f X \bar{Y_R} f_R$,
where $X$ is a scalar hidden dark matter particle, $Y_{L,R}$ are chiral fermions and $f_{L,R}$ are standard model fermions.  With such interactions hidden-sector dark matter can allow interesting direct and indirect detection phenomenology \cite{Feng:2008dz}.  For a fermionic hidden dark matter detection might be more difficult as low-energy interactions involve higher dimensional operators such as the dimension 6 four-fermion interaction $(\bar{\chi} \chi) (\bar{f} f)$.  This type of interaction  scales like $\sigma_I \sim {\lambda^4}/{\mathsf{s}}$ above an interaction scale $m_I$ and as $\sigma_I \sim ({\lambda_I^4}/{m_I^4}) \mathsf{s}$ far below it, where $\mathsf{s} \sim E_{\rm cm}^2$. Demanding that the hidden sector remain out of thermal contact with the visible sector likely requires that this effective interaction is extra-weak \cite{Sigurdson:2009uz} (weaker than the Fermi interaction strength $G_F$), but detailed predictions for nongravitational signatures of hidden-sector dark matter are best done on a case by case basis.

\section{Conclusion}
\label{sec:conclusion}

Dark matter might  belong to a hidden sector with new particle physics totally unknown to us. Even if this is the case, we can put constraints on its particle properties from its gravitational effects on the Universe. The constraints we discuss here are valid for a wide class of hidden-sector models of dark matter where the dark matter is a thermal relic particle that was once in equilibrium with a hidden-sector plasma.   We have allowed the dark-matter annihilation cross section to be a free parameter and the hidden sector temperature to be different from the visible sector. Due to these extra freedoms we have shown, in contrast to a visible-sector WIMP, we must include cases where the dark matter is relativistic, semirelativisitc, and nonrelativistic at freeze-out.   By solving the general freeze-out scenario numerically we have treated all cases in a unified way, and have found the region of hidden dark matter parameter space that can account for the observed density of dark matter while remaining consistent with current constraints on the free-streaming length and phase-space density of dark matter.   A thermal relic in a hidden sector with dark matter mass $m_{\chi} \leq 1.5~{\rm keV}$ is incompatible with current constraints.  This lower bound on the mass of hidden-sector is insensitive to the details of hidden-sector particle physics and relies on gravitationally-mediated signatures of dark matter from cosmology.
\\
\section*{Acknowledgements}
We thank Anindya Mukherjee for helpful discussions. This research is supported in part by a Natural Sciences and Engineering Research Council (NSERC) of Canada Discovery Grant.


\appendix
\section{Surface of Allowed DM Abundance}

We have determined a fitting formula which might be useful for characterizing  thermal hidden dark matter models of the kind we study here. 
We provide a polynomial fitting function for hidden dark matter scattering cross-section $\sigma_{\chi}= f( m_{\chi}, \xi)$ which gives the observed dark matter relic density. 
This formula is valid for  a general thermal warm or cold hidden dark matter model with a dark matter mass range $m_{\chi} \sim \rm{keV}$ to $100$ $ \rm{GeV}$,  range for the hidden-to-visible sector temperature ratio from $\xi \sim  0$ to $1$, and for cross-sections ranging from $10^{-3} \, \sigma_0$ to $ 10 \, \sigma_0$ where $\sigma_0 = 10^{-9} \rm{GeV}^{-2}$.
This function can be used to estimate  $\sigma_{\chi}$  for a given dark matter mass $m_{\chi}$ and  hidden-to-visible temperature ratio $\xi$. 

\begin{figure}[b]
\includegraphics[width=3.312in]{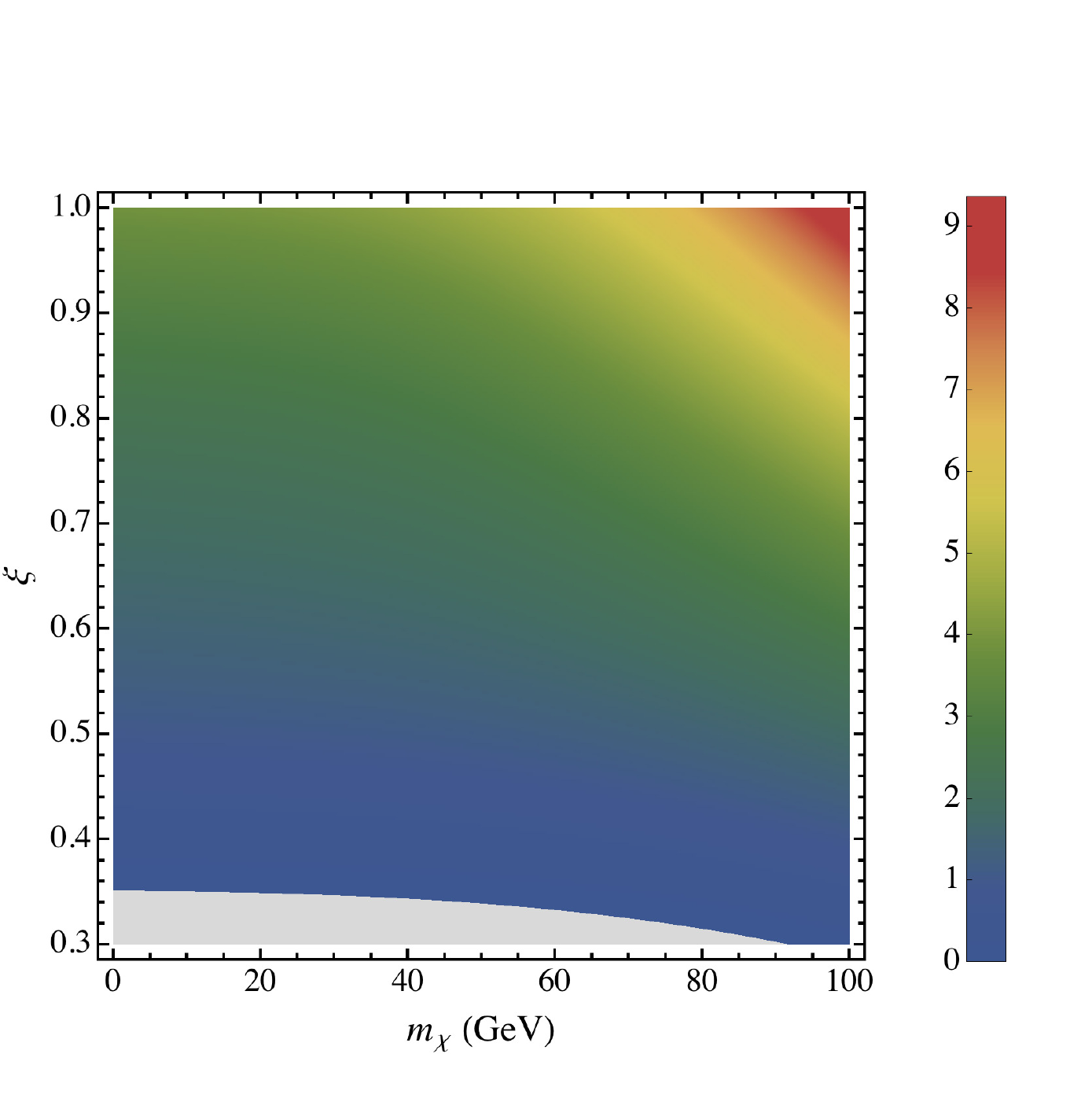}
\caption{Density plot of Eq.~(\ref{fitting}) showing $(\sigma_{\chi}/{\sigma_0})$ values in the ($\xi, m_{\chi}$) plane that yield  the allowed dark-matter relic density.  The legend  maps the value of
 $(\sigma_{\chi}/{\sigma_0})$ to an intensity or color. The grey area indicates a region where the cross section is less than $10^{-3} \,\sigma_0 $ }\label{Contourscross}
\end{figure}

\begin{align}
\sigma( m_{\chi}, \xi) = \sigma_0 \, \sum_{i,j=0}^{3} C_{ij}  \left(\frac{m_{\chi}}{{\rm GeV}}\right)^i \xi^j \, ,
\label{fitting}
\end{align}
where the mass is written in units of \rm{GeV} and $\sigma_0 = 10^{-9} \rm{GeV}^{-2}$. We have checked that the above fitting function is accurate to within 4 or 5 percent by comparing it with the actual solution of the Boltzmann
equation.
\begin{table}[h]
\begin{tabular}{|l|c|c|c|c|c|}
\hline
& $C$ & $i=0$ & $i=1$ & $i=2$ & $i=3$\\ \hline
& $j=0$ & $-5.2345$ & $-0.0016$ & $0.0001$ & $1.028 \times 10^{-6}$\\ \hline
& $j=1$ & $23.3984$ & $0.0053$ & $-0.0004$ & $-4.0929 \times 10^{-6}$\\ \hline
& $j=2$ & $-29.4988$ & $0.0029$ & $0.0005$ & $5.4513 \times 10^{-6}$\\ \hline
& $j=3$ & $15.2099$ & $-0.0018$ & $-0.0002$ & $-2.6184 \times 10^{-6}$\\ \hline
\end{tabular}
\caption{Coefficients  $C_{ij}$ of the fitting function for the hidden dark matter annihilation cross section surface with allowed relic abundance.}
\end{table}

{}

\end{document}